\documentstyle[preprint,prc,aps,floats,epsf]{revtex}

\tightenlines

\newcommand{\beq}{\begin{equation}}
\newcommand{\eeq}{\end{equation}}
\newcommand{\bea}{\begin{eqnarray}}
\newcommand{\eea}{\end{eqnarray}}
\newcommand{\ben}{\begin{eqnarray*}}
\newcommand{\een}{\end{eqnarray*}}

\newcommand{\dt}{\partial_t}

\newcommand{\kf}{k_{\rm F}}
\newcommand{\wt}{\widetilde}
\newcommand{\kt}{\widetilde k}
\newcommand{\pt}{\widetilde p}
\newcommand{\qt}{\widetilde q}
\newcommand{\wh}{\widehat}

\newcommand{\edens}{{\cal E}}
\newcommand{\order}[1]{{\cal O}(#1)}

\newcommand{\psihat}{\widehat\psi}
\newcommand{\xvec}{{\bf x}}
\newcommand{\dagphan}{{\phantom{\dagger}}}
\newcommand{\kvec}{{\bf k}}

\newcommand{\ak}{a^\dagphan_\kvec}
\newcommand{\akdag}{a^\dagger_\kvec}
\newcommand{\akv}[1]{a^\dagphan_{\kvec_{#1}}}
\newcommand{\akdagv}[1]{a^\dagger_{\kvec_{#1}}}
\newcommand{\akp}{a^\dagphan_{\kvec'}}
\newcommand{\akpdag}{a^\dagger_{\kvec'}}

\def\vec#1{{\bf #1}}

\newcommand{\nab}{\overrightarrow{\nabla}}
\newcommand{\nabsq}{\overrightarrow{\nabla}^{2}\!}
\newcommand{\nabl}{\overleftarrow{\nabla}}
\newcommand{\galnab}{\tensor{\nabla}}
\newcommand{\psid}{{\psi^\dagger}}
\newcommand{\idt}{{i\partial_t}}
\newcommand{\Sthree}{{\delta_{11'}(\delta_{22'}\delta_{33'}%
        -\delta_{23'}\delta_{32'})%
        +\delta_{12'}(\delta_{23'}\delta_{31'}-\delta_{21'}\delta_{33'})%
        +\delta_{13'}(\delta_{21'}\delta_{32'}-\delta_{22'}\delta_{31'})}}
\newcommand{\Stwo}{{\delta_{11'}\delta_{22'}-\delta_{12'}\delta_{21'}}}
\newcommand{\Left}{{\cal L}}
\newcommand{\psidagger}{\psi^\dagger}

\begin{document}
\draft

\title{Field Redefinitions at Finite Density}

\author{R.~J.~Furnstahl\thanks{{\tt furnstahl.1@osu.edu}},
        H.-W.\ Hammer\thanks{{\tt hammer@mps.ohio-state.edu}},
        and
        Negussie Tirfessa\thanks{{\tt tirfessa.2@osu.edu}}}
\address{Department of Physics \\
         The Ohio State University,\ \ Columbus, OH\ \ 43210}

\date{December 4, 2000}

\maketitle

\begin{abstract}
The apparent
dependence of nuclear matter observables on off-shell properties
of the two-nucleon potential is re-examined in the context of effective
field theory (EFT). Finite density (thermodynamic)
observables are  invariant under field redefinitions, which extends 
the well-known theorem about the invariance of S-matrix elements.
Simple examples demonstrate how field redefinitions can shift contributions
between purely off-shell two-body interactions 
and many-body forces, leaving
both scattering and finite-density observables unchanged.
If only the transformed
two-body potentials are kept, however, 
the nuclear matter binding curves will depend on the off-shell
part (generating ``Coester bands''). The correspondence between
field redefinitions and unitary transformations, which have 
traditionally been used to generate \lq\lq phase-equivalent''
nucleon-nucleon potentials, is also demonstrated.
\end{abstract}

\bigskip

\thispagestyle{empty}

\newpage

\section{Introduction}
\label{intro}

Over thirty years ago, it was observed that ``phase-equivalent'' 
nucleon-nucleon
potentials gave different predictions when used to calculate the energy
per particle of nuclear matter 
\cite{COESTER70,SRIVASTAVA70,HAFTEL71,MONAHAN71}.
Phase equivalent means that the potentials predict
identical two-nucleon observables (phase shifts and binding energies);
families of such potentials can be constructed via unitary 
transformations \cite{EKSTEIN60}.
The finite-density
discrepancies were attributed to the differing off-shell behavior
of the nucleon-nucleon (NN) amplitude or T-matrix, 
and were related semi-quantitatively to the distortion
of the two-body relative wave function in the medium
(the ``wound integral'') \cite{COESTER70}.
It was proposed that comparisons of nuclear matter calculations
could help to determine the 
correct off-shell dependence \cite{MONAHAN71,SRIVASTAVA75}. 

From the modern perspective of effective field theory (EFT), 
these results demand
a different interpretation, since off-shell effects are not 
observable.
Phase-equivalent EFT potentials should give the same result for 
{\em all\/} observables (up to
expected truncation errors) 
if consistent power counting is used; differences
imply that something is missing.
In this paper, we show how this issue is resolved
by considering field redefinitions at finite density
in an effective field theory.

There is an extensive literature, primarily from the sixties and seventies,
on the role of off-shell physics in nuclear phenomena 
(see, e.g., Ref.~\cite{SRIVASTAVA75} and references therein).  
This includes
not only few-body systems (e.g., the triton) and nuclear matter, 
but interactions of two-body systems with external probes,
such as nucleon-nucleon bremstrahlung and the electromagnetic form
factors of the deuteron.
The implicit 
premise, which persists today, is that there is
a true underlying potential governing the nucleon-nucleon force,
so that its off-shell properties can be determined.
Indeed,
the nuclear many-body problem has traditionally been posed as 
finding approximate solutions to the many-particle Schr\"odinger
equation, given a fundamental two-body interaction that reproduces
two-nucleon observables.

An alternative approach to nuclear phenomena is that of effective
field theory.
The EFT approach 
exploits the separation of scales in physical 
systems \cite{LEPAGE89,KAPLAN95,GEORGI94,EFT98,EFT99,Birareview,BEANE99}.
Only low-energy (or long-range) degrees of freedom are included
explicitly, with the rest parametrized in terms of the most general
local (contact) interactions.  Using renormalization,
the influence of high-energy states on low-energy observables
is captured in a small number of constants.
Thus, the EFT describes universal low-energy physics independent of
detailed assumptions about the high-energy dynamics.
The application of EFT methods to many-body problems promises 
a consistent organization of many-body corrections, with reliable error
estimates, and insight into the analytic structure of observables.
EFT provides a model independent description of finite density
observables in terms of parameters that can be fixed from
scattering in the vacuum.

A fundamental theorem of quantum field theory states that
physical observables (or more precisely, S-matrix elements) are
independent of the choice of interpolating fields that appear in
a Lagrangian \cite{HAAG58,COLEMAN69}.
Equivalently, observables are invariant 
under a change
of field variables in a field theory Lagrangian (or Hamiltonian).
This ``equivalence theorem'' holds for renormalized field theories, although
there are subtleties with some regularizations \cite{BERGERE76}. 
In an EFT, one exploits the invariance under field redefinitions
to eliminate redundant terms in the effective Lagrangian and to choose
the most convenient or efficient form for practical 
calculations \cite{POLITZER80,GEORGI91,KILIAN94,SCHERER95a,ARZT95,Man98}.
Since off-shell Green's functions and the corresponding off-shell amplitudes
{\it do\/} change under field redefinitions, one must conclude that off-shell
properties are unobservable.

Several recent works have emphasized from a field theory 
point of view the impossibility of observing off-shell effects.
In Refs.~\cite{FEARING98,FEARING99}, 
model calculations are used to illustrate how apparent determinations of
the two-nucleon off-shell T-matrix in  
nucleon-nucleon bremstrahlung are illusory, since field redefinitions
shift contributions between off-shell contributions and contact interactions
(see Sec.~\ref{examples}). 
Similarly, it was shown in Ref. \cite{SCHERER95b} that Compton scattering
on a pion cannot be used to extract information on the off-shell
behavior of the pion form factor.
The authors of Refs.~\cite{FHK99,CFM96} emphasized the nonuniqueness
of chiral Lagrangians for three-nucleon forces and pion production.
Field redefinitions lead to different off-shell forms that yield the same
observables within a consistent power counting.
In Ref.~\cite{KSW99}, an interaction proportional to the equation of
motion is shown to have no observable consequence
for the deuteron electromagnetic form factor, even though it contributes
to the off-shell T-matrix.

In systems with more than two nucleons, one can trade off-shell, two-body
interactions for many-body forces.
This explains how two-body interactions related by unitary transformations 
can predict different binding energies for the triton \cite{ARNAN73}
if many-body forces are not consistently included.
These issues are discussed from the viewpoint of unitary transformations
in Refs.~\cite{POLYZOU90} and \cite{AMGHAR95}.
An EFT analysis of the triton with short-range forces was performed
by Bedaque and collaborators.
They showed that a three-body counterterm is needed in the triton
for consistency \cite{BEDAQUE98}. 
Varying this counterterm generates a one-parameter family of observables,
which explains the Phillips line.

Here we consider fermion systems in the thermodynamic limit.
We extend the equivalence theorem to finite density observables,
or thermodynamic observables in general, and show the consequence of
field redefinitions.
Since the thermodynamic observables are conveniently expressed
in a functional integral formalism,
we adapt
the discussion in Ref.~\cite{ARZT95}, which provides a particularly transparent
(if not entirely rigorous) account of the equivalence theorem
for S-matrix elements
and how it can be applied in EFT's, based on functional integrals.
We show how field redefinitions applied to an EFT Lagrangian with only
on-shell vertices generate new off-shell vertices, which by themselves
change predictions for an infinite system.  However, many-body vertices
are also induced, and if included consistently, will precisely cancel the
off-shell parts.
In the end, all calculations agree, with no off-shell ambiguities.

In section II,  we review the argument 
in Ref.~\cite{ARZT95}, extend it to thermodynamic
observables, and show how field redefinitions can reduce an
effective Lagrangian
to a canonical form.
In section III, we illustrate the effects of field redefinitions with 
a simple example applied to the dilute Fermi gas of Ref.~\cite{HAMMER00}.
The connection to traditional treatments is elucidated in section IV.
Finally, we summarize in Section V. 


\section{Field Redefinitions}
\label{FLDRDF}

The equivalence theorem is typically stated as the invariance
of the S-matrix for asymptotic scattering states under
a change of field variables.
Intuitive diagrammatic proofs, as in Refs.~\cite{COLEMAN88,BANDO88},
are based on the reduction formula for calculating S-matrix elements
from on-shell Green's functions.
The proofs show that 
field redefinitions change only the wave function
renormalization, which is divided out in the reduction formula.
The extension to observables in the thermodynamic limit, such as the
energy density of an infinite system of fermions at zero temperature,
is not obvious since there are no asymptotic single-particle states.
Here we make that extension for nonrelativistic
systems and discuss the consequences for effective field theories.
The generalization to covariant effective field theories is
straightforward.

A rigorous proof of the equivalence theorem must include a careful discussion
of renormalized Green's functions and amplitudes, 
and the impact of different regulators.
Such a proof, based on the renormalized quantum action principles, is given
in Ref.~\cite{BERGERE76} and extended to a heavy-fermion expansion (in the
context of heavy quark effective field theory) in Ref.~\cite{KILIAN94}.
The new issues in considering thermodynamic observables, however, do not
change the subtle technical issues in the proof, 
so we build instead on a less rigorous
but more intuitive path integral discussion of the equivalence
theorem \cite{ARZT95}
to illustrate the broad concepts and consequences.
The path-integral formalism is convenient, since one
can generate Green's functions and subsequently S-matrix elements,
as well as thermodynamic observables, in the same framework.

We will only consider heavy (nonrelativistic) boson or
fermion fields explicitly;
the generalization
to include long-range bosonic fields (e.g., such as the pion) is 
straightforward.
The procedure that leads from an underlying 
relativistic theory to an effective Lagrangian ${\cal L}$
in a heavy-particle formulation  is outlined in
Ref.~\cite{Birareview}.
We work in the rest frame, so there are no 
non-trivial four-velocities associated with the field variables.
In general, ${\cal L}$ contains all terms consistent with the usual
space-time symmetries as well as any symmetries of the
underlying theory.
These include, in general, higher-order time derivatives, which can
be eliminated as described below. 
The path integral form of the generating functional is
written  schematically as:
\beq
  Z[\xi^\dagger,\xi] = \int\! {\cal D}(\psidagger,\psi)\,
     e^{i\int{\cal L}[\psi,\psidagger] + \xi^\dagger \psi
        + \psidagger \xi} \ ,    
\eeq
where $\psi$ and $\psi^\dagger$ are Grassmann variables for fermions,
$\xi^\dagger$ and $\xi$ are external Grassmann sources.
By taking functional derivatives of $Z$
with respect to the sources, we generate
the $n$-body Green's functions and then S-matrix elements
via a reduction formula.

Here we follow the discussion of the equivalence theorem
in Ref.~\cite{ARZT95}.
The basic observation is that 
if we change field variables in the functional integral, 
predictions for given external sources should be unchanged.
Consider a transformation of $\psi$ of the form:
\beq
  \psi(x) \longrightarrow \psi(x) + \eta T[\psi,\psidagger]
    \ ,  \label{eq:psitrans}
\eeq
where $T[\psi,\psidagger]$ is a local polynomial in $\psi(x)$, 
$\psidagger(x)$, and their derivatives, and $\eta$ is an
arbitrary counting parameter (to be used below).
The transformation given by Eq.~(\ref{eq:psitrans}) is local and respects
the symmetries of ${\cal L}$. There are no space-time dependent external
parameters, i.e.\ the $x$-dependence of $T[\psi,\psidagger]$ comes
from the fields alone. Specific examples of such transformations 
are given in Eqs.~(\ref{ftrafo}) and (\ref{eq:psitran}).

Within the path integral, three types of change result 
from Eq.~(\ref{eq:psitrans}):
\begin{enumerate}
  \item the Lagrangian changes: 
   ${\cal L} \rightarrow {\cal L}'$;
  \item there is a Jacobian from the transformation;
  \item there are additional terms coupled to the external sources
   $\xi$ and $\xi^\dagger$.
\end{enumerate}
{\em The statement of the equivalence theorem is that changes 2 and 3 
by themselves do
not affect observables, so that one is free to change variables in
the Lagrangian alone.\/}
By introducing ghost fields,
the Jacobian from 2 can be exponentiated.
As detailed in Ref.~\cite{ARZT95}, new terms involving ghosts
either decouple or contribute pure power divergences in ghost loops. 
The latter are either zero in dimensional regularization or
can be absorbed without physical consequence into couplings
in ${\cal L}$.
Therefore the Jacobian has no observable effect.%
\footnote{Note that the field redefinitions Eq.~(\ref{eq:psitrans})
are different from the transformations used to explore anomalies,
where contributions from the Jacobian {\em are\/} critical \cite{DONOGHUE92}.
The latter transformations involve space-time dependent external
parameters.}
The extra terms in 3 do change the Green's functions but not the
S-matrix.
The proof is analogous to the usual diagrammatic proof \cite{ARZT95}.

To extend the discussion to finite density and/or temperature, we 
consider a Euclidean path integral formulation of the partition function $Z$
and thermodynamic potential $\Omega$:
\beq
  Z \equiv \exp(-\beta\Omega) \equiv
  {\rm Tr} \{ \exp[-\beta(\widehat H - \mu \widehat N)] \} 
  \ ,
  \label{eq:Omega}
\eeq
where $\beta$ is the inverse temperature, $\mu$ is the chemical potential,
$\widehat H$ is the Hamiltonian, 
and $\widehat N$ is the number
operator (e.g., the baryon number for nuclear systems).
The hat indicates second quantized operators that act in Fock space. 
We direct the reader to 
Ref.~\cite{NEGELE88} for details on the derivation and interpretation
of nonrelativistic path integrals at finite temperature and chemical 
potential.
The connection between $\widehat H$ and
${\cal L}$ is complicated, but we do not need to actually
derive the connection, since we assume a general form for ${\cal L}$. 
However, we need to be more explicit with $\widehat N$, which
is not in general equal to the spatial integral
of $\psidagger\psi$.  Instead, the appropriate operator should be derived
using Noether's theorem applied to ${\cal L}$.

The integrations in the path integral corresponding to Eq.~(\ref{eq:Omega}) 
will be over complex variables satisfying periodic boundary
conditions in imaginary time 
for bosons and over Grassmann variables satisfying antiperiodic
boundary conditions for fermions.
If there is no chemical potential,
the extension of the equivalence theorem to finite temperature is
immediate, because the finite temperature boundary conditions
are trivially preserved by field redefinitions of the form
in Eq.~(\ref{eq:psitrans}) and the remainder of the argument is unchanged.
Thus the thermodynamic quantities derived from the partition function
(e.g., free energies and specific heats)
will be invariant under field redefinitions for all $\beta$.
Since there are no external sources, we are free to introduce 
linearly coupled
sources to construct a perturbative expansion {\em after\/} the field
redefinitions.

A nonzero chemical potential will be multiplied
by new terms after the field redefinition.
The conserved current is associated with a global $U(1)$ symmetry 
transformation 
\beq
   \psi(x) \longrightarrow e^{-i\phi}\psi(x) \quad \mbox{ and } \quad
   \psid(x) \longrightarrow e^{i\phi}\psid(x)\ ,
\eeq
under which ${\cal L}$ is invariant.
We can conveniently identify the Noether current by promoting $\phi$
to a function of $x$ and considering infinitesimal 
transformations with \cite{DONOGHUE92}
\beq
  {\cal L} \longrightarrow \widetilde{\cal L}[\psi,\psidagger;\phi(x)]
   \ .
\eeq
Then the charge density $\widehat\rho$ is given by
\beq
  \widehat\rho \equiv  \frac{\delta}{\delta(\partial_t \phi)}
           \widetilde{\cal L}[\psi,\psidagger;\phi(x)] 
           \equiv F[\psi,\psidagger] \ .
\eeq
The possibility of higher-order time derivatives or time derivatives
in non-quadratic terms in $\widetilde{\cal L}$ means that, in general,
$F[\psi,\psidagger] \neq \psidagger\psi$.

Under the transformation Eq.~(\ref{eq:psitrans}),
\bea
  {\cal L}[\psi,\psidagger] 
     \longrightarrow {\cal L}'[\psi,\psidagger] 
     &=&  {\cal L}[\psi + \eta T,\psidagger+\eta T^\dagger]
     \nonumber \\
     &=&  {\cal L} + \Bigl(\eta T 
         \frac{\delta {\cal L}}{\delta\psi} + \mbox{H.c.} \Bigr)
         + \cdots \ ,
     \label{eq:calmu}
\eea
while 
\bea
  F[\psi,\psidagger] \longrightarrow 
  F'[\psi,\psidagger] &=&
  F[\psi + \eta T,\psidagger+\eta T^\dagger]
     \nonumber \\
    &=& F + \Bigl(\eta T 
         \frac{\delta F}{\delta\psi} + \mbox{H.c.} \Bigr)
         + \cdots \ , 
    \label{eq:Fnew}
\eea
where we have written the first terms in the
functional Taylor expansions of Eqs.~(\ref{eq:calmu}, \ref{eq:Fnew})
and H.c.\ denotes the Hermitian conjugate.
If $T$ contains no time derivatives (as in the examples in
Sections \ref{examples} and \ref{unitary}), then
using the fact that $\delta/\delta(\partial_t\phi)$
commutes with $\delta/\delta\psi$, we see order-by-order in $\eta$ that
\beq
   {\wh\rho}\,' \equiv \frac{\delta}{\delta(\partial_t \phi)}
           \widetilde{\cal L}' [\psi,\psidagger;\phi(x)] 
        =    F'[\psi,\psidagger] \ . 
\eeq
So the term multiplying $\mu$ in the transformed path integral
{\em is\/} the new charge density.
If $T$ contains time derivatives, ${\wh\rho}\,'$ will contain additional 
terms proportional to the equations of motion, which give zero
contribution \cite{KILIAN94,ARZT95}.

At zero temperature, we can work without an explicit chemical potential, 
but instead specify the density
through boundary conditions in the noninteracting propagator
(see Sect.~\ref{examples}).
Using the new charge density operator, we find the density is unchanged
by the transformation.  This is illustrated explicitly for a
special case in the Appendix.

For an effective field theory, invariance under nonlinear field redefinitions
has practical and philosophical implications.
On the practical side, because the EFT already contains all terms consistent
with the underlying symmetries (which are preserved by the field variable
changes we consider), there is no penalty for making such transformations:
they just change the coefficients of existing terms.  Therefore, we can
use this freedom to remove redundant terms and define a canonical form of
the EFT Lagrangian.
This can simplify the calculation or improve its convergence.
One possibility is what Georgi calls ``on-shell effective field theory,''
which features only those vertices that are non-zero for on-shell kinematics
(of free particles) \cite{GEORGI91}.
This was the choice in Ref.~\cite{HAMMER00}, which treated dilute Fermi
systems.

An effective Lagrangian is organized according to a hierarchy,
which we
label by $\eta$ \cite{ARZT95}:
\beq
  \Left = \sum_{n=0}^{\infty} \eta^n \Left_n 
     \ .  \label{eq:Lsum}
\eeq
For example, we have $\eta \propto 1/\Lambda$
for a natural field theory as in Ref.~\cite{HAMMER00}.
Define $\Left_0$ to be:
\beq
  \Left_0 = \psidagger \biggl[ i\partial_t
         + \frac{\nabsq}{2M} \biggr] \psi
         \ .  \label{eq:Lzero}
\eeq
As a result of the transformation Eq.~(\ref{eq:psitrans}), 
$\Left \rightarrow \Left'$, with
\beq
  \Left' = \Left_0 + \biggl(\eta T[\psi,\psidagger]
         \biggl[ i\partial_t
         + \frac{\nabsq}{2M} \biggr] \psi + \mbox{H.c.} \biggr)
       + \sum_{n=1}^{\infty} \eta^n \Left'_n
       \ . \label{eq:Lprime} 
\eeq
Thus, at leading order in $\eta$, we generate a term proportional
to the free equation of motion.
The corresponding interaction term will be proportional to the
inverse free propagator, and thus will vanish on shell.
So we can change 
off-shell Green's functions (e.g., the two-nucleon
off-shell T-matrix) 
by field transformations without changing observables.
The consequence is that one can change the appearance of ${\cal L}$.
One may find that better convergence of an EFT expansion
is correlated with a particular off-shell
dependence, but {\em the idea of measuring the ``correct'' off-shell 
behavior has to be dismissed}.
These implications are extended to conventional approaches
to nuclear phenomena by considering them as implementations, albeit 
incomplete, of the EFT approach. The discussion can be directly
extended to a finite system with a single particle potential
included in ${\cal L}_0$ (and in the noninteracting propagator).

We are also able to remove redundant terms,
which was implicit in the choice of the ``general'' effective Lagrangian
in Ref.~\cite{HAMMER00}.
Time derivatives other than the quadratic term can be systematically 
eliminated by field redefinitions of the form in Eq.~(\ref{eq:psitrans})
by exploiting the EFT hierarchy.
In particular, the lower-order (in $\eta$) terms can be removed at the
cost of changing the coefficients of higher-order terms in $\eta$. 
For example, suppose we have the Lagrangian
\beq
  {\cal L} = \psidagger \biggl[ i\partial_t
         + \frac{\nabsq}{2M} \biggr] \psi
         + \eta_1 \psidagger(i\partial_t)^2 \psi
         + \eta_2 \bigg(\psidagger \frac{\galnab^2}{2M} (i\partial_t) \psi
                  +\mbox{ H.c.} \bigg)
         + {\cal O}(\eta^2) \ ,
\eeq
where $\galnab=\nabl-\nab$. The field redefinition
\bea
  \psi &\longrightarrow& \psi
    - \frac{\eta_1}{2} (i\partial_t\psi)
    + \Bigl(\frac{\eta_1}{2} - 4 \eta_2 \Bigr) \frac{\nabsq}{2M}\psi\ ,
  \nonumber\\
  \psid &\longrightarrow& \psid
    + \frac{\eta_1}{2} (i\partial_t\psid)
    + \Bigl(\frac{\eta_1}{2} - 4 \eta_2 \Bigr) \frac{\nabsq}{2M}\psid\ ,
\eea
results in the new Lagrangian
\beq
  {\cal L}' = \psidagger \biggl[ i\partial_t
         + \frac{\nabsq}{2M} \biggr] \psi
         + \Bigl(\eta_1 - 8 \eta_2 \Bigr)
           \psidagger \Bigl(\frac{\nabsq}{2M}\Bigr)^2 \psi
         + {\cal O}(\eta^2) \ ,
\eeq
with no time derivatives outside of ${\cal L}_0$ to order $\eta$.
Once we remove terms up to some order in $\eta$, further transformations
will not reintroduce them because they involve the next higher order.%
\footnote{For another example of such a field redefinition see 
Ref. \cite{BeG00}.}

Equation~(\ref{eq:Lprime}) shows 
that different choices of effective Lagrangians, related
by field redefinitions, can differ by off-shell contributions.
The equivalence theorem guarantees that these must 
ultimately be canceled in observables (or vanish individually).
In a sense, that makes any explicit demonstration ``trivial''.
The realization of the equivalence theorem, however, is often
counterintuitive. Approximations in a truncated theory have to be 
consistent with this realization.
A common source of confusion arises in discussions that assume
only two-body vertices are kept in many-body calculations.
Therefore, we turn to some concrete examples.


\section{Perturbative Examples}
\label{examples}
\subsection{Transformation of the Lagrangian}
In this section, we illustrate the connection between off-shell 
two-body terms and
three-body terms using a simple field redefinition. In particular, the
effect of an off-shell vertex generated by a field redefinition is 
exactly canceled by a three-body term generated by the same transformation.
We demonstrate this cancellation explicitly for the process of $3\to 3$
scattering in the vacuum and for the energy density of a dilute Fermi system
at finite density.
For simplicity, we consider a natural EFT for heavy nonrelativistic
fermions of mass $M$ that has spin independent $s$-wave interactions 
whose strength is correctly estimated by naive dimensional analysis.
Such a theory is discussed in detail, e.g., in 
Refs.~\cite{HAMMER00,Kaplantalk}.

For a natural theory all effective range parameters are of order
$1/\Lambda$, where $\Lambda$ is the breakdown scale of the theory.
For momenta $k\ll\Lambda$, all interactions appear short-ranged and
can be modeled by contact terms. We consider
a local Lagrangian for a nonrelativistic fermion
field that is invariant under Galilean, parity, and time-reversal 
transformations:
\bea
  {\cal L}  &=&
       \psi^\dagger \biggl[i\partial_t + \frac{\nabsq}{2M}\biggr]
                 \psi - \frac{C_0}{2}(\psi^\dagger \psi)^2
            + \frac{C_2}{16}\Bigl[ (\psi\psi)^\dagger 
                                  (\psi\galnab^2\psi)+\mbox{ H.c.} 
                             \Bigr]  +  \ldots 
\,,\label{lag}
\eea
where $\galnab=\overleftarrow{\nabla}-\nab$ is the Galilean invariant
derivative and H.c.\ denotes the Hermitian conjugate. 
Higher-order time derivatives are not included, since it is most
convenient for our purposes to eliminate them in favor of gradients,
as described in Sec.~\ref{FLDRDF}.
Naive dimensional 
analysis estimates $C_{2i}$ to be roughly
$4\pi/(M\Lambda^{2i+1})$, which implies a
perturbative expansion in $k/\Lambda$. $C_0$ and $C_2$
can be related to the $s$-wave scattering length $a$ and effective
range $r_e$ via a matching calculation. This leads to the relations
\beq
      C_0 = \frac{4\pi a}{M} \quad \mbox{and}\quad C_2=C_0 \frac{a r_e}{2}\,.
      \label{C2imatch} 
\eeq         

We stress that we choose this simple example to make the cancellation
of off-shell effects in observables as transparent
as possible. As follows from the general discussion in the previous section,
analogous cancellations occur in more complicated theories 
that include long-range physics, relativistic effects, spin-dependent 
interactions, nonperturbative resummations and so on.

We generate a new Lagrangian ${\cal L}'$ by performing the field 
transformation
\bea
\psi &\to& \psi+\frac{4\pi\alpha}{\Lambda^3}(\psi^\dagger \psi)\psi\,,
    \qquad
    \psi^\dagger \to \psi^\dagger+\frac{4\pi\alpha}{\Lambda^3}\psi^\dagger 
    (\psi^\dagger \psi)\,,
       \label{ftrafo}
\eea
on ${\cal L}$. The factor $1/\Lambda^3$ is introduced  to keep 
the arbitrary parameter $\alpha$ dimensionless. The additional factor
of $4\pi$ is introduced for convenience. With this choice,
the induced terms have the size expected from naive dimensional analysis 
if $\alpha$ is of order unity. We obtain for ${\cal L}'$:
\bea
{\cal L}'&=&{\cal L}-\frac{4\pi\alpha}{\Lambda^3}2 C_0 (\psid\psi)^3
      \nonumber\\
 &+&\frac{4\pi\alpha}{\Lambda^3}\frac{C_2}{8}\left
    \{2(\psid\psi)(\psi\psi)^\dagger
   (\psi\galnab^2\psi)+(\psid\psi)^2 \left[(\nabsq \psid)\psi+
    \psid(\nabsq \psi)
    +2(\nab \psid)(\nab \psi)\right]+\mbox{ H.c.} \right\}
      \nonumber\\
 &+&\frac{4\pi\alpha}{\Lambda^3}\left\{(\psid\psi)\psid(\idt\psi)
   -\frac{1}{2M}\left[\psid(\nab\psi)\cdot\psid(\nab\psi)+2(\nab
   \psid)\psi\cdot\psid(\nab\psi)\right]+\mbox{ H.c.} \right\}
      \nonumber\\
 &+& \mbox{ 4- and higher-body terms } + {\cal O}(\alpha^2)\,,
\eea
where 4- and higher-body terms as well as terms of ${\cal O}(\alpha^2)$
have been omitted.

\begin{figure}[t]
 \epsfclipon 
 \epsfxsize=15.cm
 \centerline{\epsffile{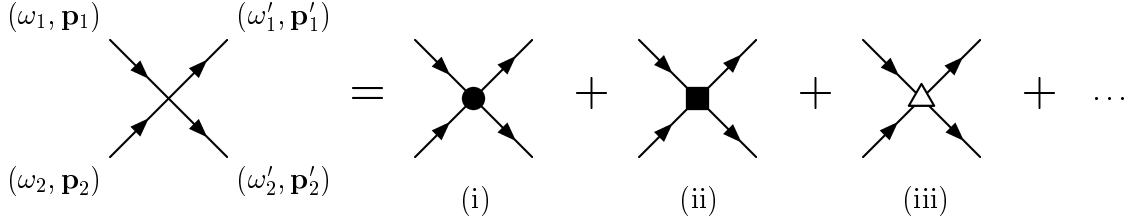}}
\vspace*{0.1in}
\caption{Two-body vertices generated by ${\cal L}'$. 
The corresponding Feynman rules are given in Eq.~(\ref{eftv2}).}
\label{eftvertex2}
\end{figure}

\begin{figure}[t]
 \epsfclipon 
 \epsfxsize=13.cm
 \centerline{\epsffile{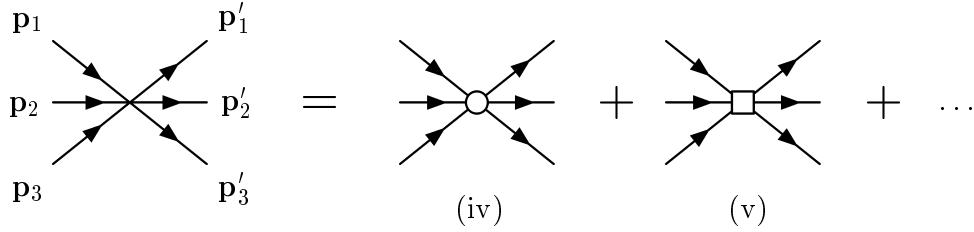}}
\vspace*{0.1in}
\caption{Three-body vertices generated by ${\cal L}'$.
The corresponding Feynman rules are given in Eq.~(\ref{eftv3}).}
\label{eftvertex3}
\end{figure}

The two-body vertices generated by the transformed Lagrangian ${\cal L}'$ 
are shown in Fig.~\ref{eftvertex2}. 
The vertices (i) and (ii) were already
present in the untransformed Lagrangian ${\cal L}$, while (iii) is a new
energy-dependent off-shell vertex of ${\cal O}(\alpha)$,
which vanishes identically if all lines are on-shell.
However, ${\cal L}'$ also contains two new three-body vertices 
of ${\cal O}(\alpha C_0)$ and  ${\cal O}(\alpha C_2)$.%
\footnote{These vertices are already present in a general version
of ${\cal L}$, in which case the new contributions change the
corresponding coefficients.  The cancellation discussed here is unchanged.} 
These vertices are illustrated in Fig.~\ref{eftvertex3}.
The spin projection of a particle with energy $\omega_i (\omega'_j)$ 
and momentum $\vec{p}_i (\vec{p'}_j)$
is denoted by $s^{\vphantom{x}}_i (s'_j)$; we abbreviate 
$\delta_{s^{\vphantom{x}}_i s'_j} \equiv \delta_{i j'}$. 
Note that $\omega_i \not= \vec{p}_i^2 /(2M)$ unless the particle is on shell.
The Feynman rules for all vertices are listed below.
\begin{itemize}
\item Two-body vertices:
\bea
\mbox{(i)} &\quad & -iC_0\,S_2
      \nonumber\\
\mbox{(ii)} &\quad & -i\frac{C_2}{8}\left\{(\vec{p}_1-\vec{p}_2)^2+(\vec{p'}_1
   -\vec{p'}_2)^2\right\}S_2
      \nonumber\\
\mbox{(iii)} &\quad & i\frac{4\pi\alpha}{\Lambda^3}\left\{
   \omega_1+\omega_2+\omega'_{1}+\omega'_{2}-\frac{1}{2M}\left[\vec{p}_1^2+
  \vec{p}_2^2+\vec{p'}_1^2+\vec{p'}_2^2\right]\right\}S_2
      \label{eftv2}
\eea
\item Three-body vertices:
\bea
\mbox{(iv)} &\quad & -i \frac{4\pi\alpha}{\Lambda^3} 12 C_0\,S_3
      \nonumber\\
\mbox{(v)} &\quad & -i \frac{4\pi\alpha}{\Lambda^3}\frac{C_2}{2}
   \sum_{{\rm Perm}\{1',2',3'\}}\delta_{11'}\delta_{22'}\delta_{33'}
     \;\epsilon_{1'2'3'} \left[ (\vec{p}_1-\vec{p}_2)^2 \right.
         \nonumber\\
     &\quad &\  \left. +(\vec{p}_1-\vec{p}_3)^2 
                          +(\vec{p}_1-\vec{p'}_1)^2 +(\vec{p}_2-\vec{p}_3)^2
                          +(\vec{p}_2-\vec{p'}_2)^2
     \right.\nonumber \\
&\quad &\  \left. +(\vec{p}_3-\vec{p'}_3)^2 +(\vec{p'}_1-\vec{p'}_2)^2
                +(\vec{p'}_1-\vec{p'}_3)^2 +(\vec{p'}_2-\vec{p'}_3)^2\right]
       \label{eftv3}
\eea
\end{itemize}
The sum in (v) runs over all permutations of $\{1',2',3'\}$
and $\epsilon_{ijk}$ is the totally antisymmetric Levi-Civita tensor
with $\epsilon_{1'2'3'}=-\epsilon_{2'1'3'}=1$.
Note that vertex (v) can be rewritten in terms of the total
momentum $\vec{P}=\vec{p}_1+\vec{p}_2+\vec{p}_3$ as
\bea
\mbox{(v)} &\quad & -i \frac{4\pi\alpha}{\Lambda^3} C_2
   \sum_{{\rm Perm}\{1',2',3'\}}\delta_{11'}\delta_{22'}\delta_{33'}
     \;\epsilon_{1'2'3'} \left[ 3 \vec{P}^2 -\vec{p}_1 \cdot\vec{p'}_1
         -\vec{p}_2 \cdot\vec{p'}_2-\vec{p}_3 \cdot\vec{p'}_3 \right]\ .
\eea
The spin functions $S_2$ and $S_3$ are given by
\bea
   S_2 &=& \Stwo\ , 
      \nonumber\\
   S_3 &=& \Sthree\ ,
\eea
for spin-1/2 fermions and by $S_2=2$ and $S_3=6$ for spin-0 bosons.
The propagator for a particle line with energy $\omega$ and momentum
$\vec{q}$ is simply
\beq
  iS_0(\omega,\vec{q})=\frac{i}{\omega-\vec{q}^2/(2M)+i\epsilon}\ .
    \label{freeprop}
\eeq

As discussed in Sec.~\ref{FLDRDF},
the Lagrangian ${\cal L}'$ has to give the same results for observables
as the original Lagrangian ${\cal L}$. For this to happen, certain 
combinations of diagrams involving vertices (iii), (iv), 
and (v) have to cancel.
As $\alpha$, $C_0$, and $C_2$ are independent parameters, this has
to occur at every order in $\alpha$, $C_0$, and $C_2$ independently.
In the following two subsections, we illustrate this cancellation for $3\to 3$
scattering in the vacuum and for the energy density at finite density.

\subsection{Cancellations at Zero Density}
As mentioned above, all diagrams within one order of $\alpha$ have to cancel 
identically. Furthermore, the power counting of the original, untransformed
theory carries over. The simplest example is the tree-level $2\to 2$
diagram with vertex (iii) which vanishes identically when all external legs 
are on shell so that the cancellation at ${\cal O}(\alpha)$ is trivial.
Similarly, it is straightforward to show that all diagrams
at ${\cal O}(\alpha C_0^k C_2^l)$ that are bubble chains including
$k$ vertices of type (i) and $l$ vertices of type (ii) plus one vertex 
of type (iii) vanish as well.
This is most easily seen in dimensional regularization, where all 
vertices are on shell. 

\begin{figure}[t]
  \epsfclipon
  \epsfxsize=13.cm  
  \centerline{\epsffile{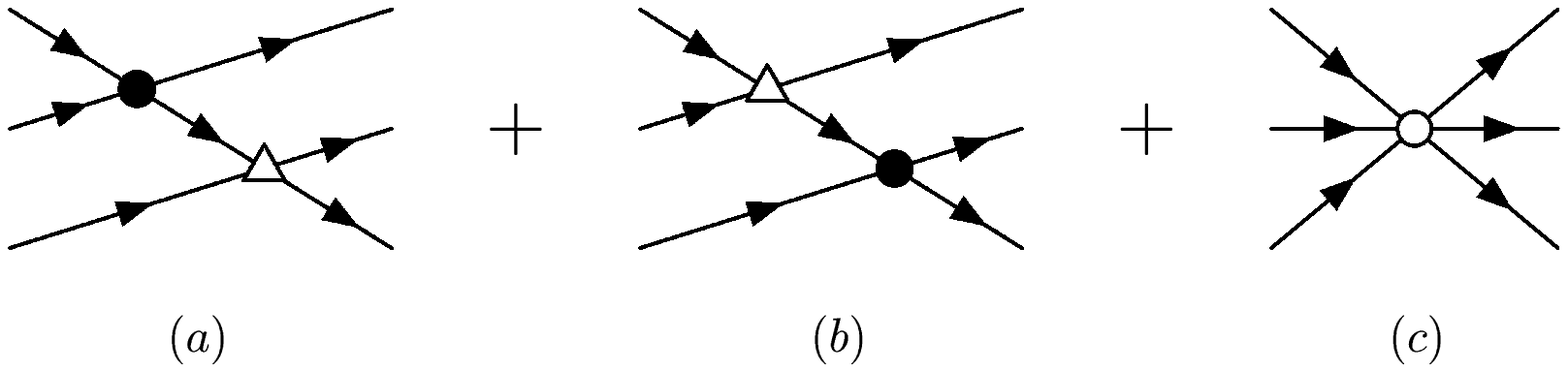}}
\smallskip
\caption{Additional diagrams for $3\to 3$ scattering generated
by ${\cal L}'$ at ${\cal O}(\alpha C_0)$. The external lines are 
labeled as in Fig. \ref{eftvertex3}.
}
\label{alphaC0}
\end{figure}

\begin{figure}[t]
  \epsfclipon
  \epsfxsize=13.cm
  \centerline{\epsffile{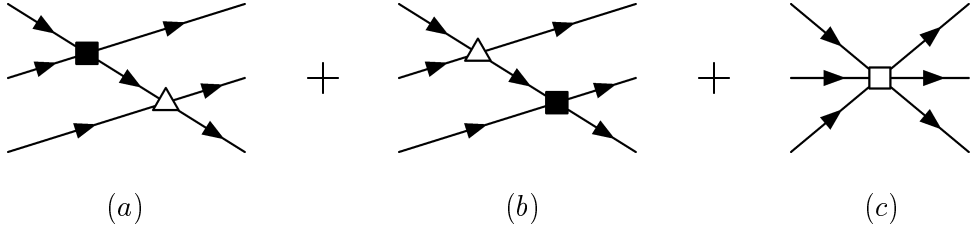}}
\smallskip
\caption{Additional diagrams for $3\to 3$ scattering generated
by ${\cal L}'$ at ${\cal O}(\alpha C_2)$. 
The external lines are labeled as in Fig. \ref{eftvertex3}.
}
\label{alphaC2}
\end{figure}

In the following, we consider $3\to3$ scattering, where 
cancellations between different diagrams occur. We concentrate on the
case of spin-1/2 fermions. The case of spin-0 bosons is analogous
but somewhat simpler, because the spin function is trivial.
At ${\cal O}(\alpha C_0)$, we have the three tree-diagrams shown in
Fig.~\ref{alphaC0}. The external lines are labeled as in Fig. \ref{eftvertex3}.
Using the Feynman rules from Eqs.~(\ref{eftv2})--(\ref{freeprop}), 
we find for  diagram \ref{alphaC0}(a),
\bea
  {{\cal M}}_{\rm\ref{alphaC0}(a)}
   &=&  i\frac{4\pi\alpha}{\Lambda^3}C_0
     \left(\delta_{11'}\delta_{22'}\delta_{33'}-\delta_{11'}\delta_{23'}
            \delta_{32'}-\delta_{12'}\delta_{21'}\delta_{33'}
     +  \delta_{13'}\delta_{21'}\delta_{32'}\right)\,,
\eea
where the internal propagator was canceled 
against the corresponding dependence 
of the off-shell vertex. At this point, we have to take into account 
that the cyclic permutations of the incoming and outgoing lines in diagram 
\ref{alphaC0}(a) generate topologically distinct diagrams that have to be 
included as well. Consequently, diagram \ref{alphaC0}(a) really represents
nine diagrams, which when taken together generate the full 
spin function $S_3$. After summing over the cyclic permutations 
of the incoming and outgoing lines, we obtain
\beq
  {\cal M}_{\rm\ref{alphaC0}(a)}
     = 6i\frac{4\pi\alpha}{\Lambda^3}C_0 S_3
     =  {\cal M}_{\rm\ref{alphaC0}(b)}   \,.
\eeq
Evaluating diagram \ref{alphaC0}(b) gives exactly the same result
as diagram \ref{alphaC0}(a). Finally, diagram \ref{alphaC0}(c) is 
simply given by vertex (iv).
Adding the contribution of all diagrams at ${\cal O}(\alpha C_0)$
we find 
\beq
  {\cal M}_{\rm\ref{alphaC0}(a)}+{\cal M}_{\rm\ref{alphaC0}(b)}+
  {\cal M}_{\rm\ref{alphaC0}(c)} = 0 \ ,
\eeq
and to ${\cal O}(\alpha C_0)$ the off-shell dependence introduced by 
vertex (iii) cancels identically.
A similar cancellation occurs for the diagrams at ${\cal O}(\alpha C_2)$
where the $C_0$ vertex in Fig.~\ref{alphaC0} is replaced by a $C_2$ vertex
and vertex (iv) is replaced by vertex (v).
The corresponding diagrams are shown in Fig.~\ref{alphaC2}.
Due to the momentum dependence of the vertices, the calculation is
somewhat more tedious but straightforward. After summing over the cyclic 
permutations of the external lines, we find
\beq
  {\cal M}_{\rm\ref{alphaC2}(a)}+{\cal M}_{\rm\ref{alphaC2}(b)}=
 -{\cal M}_{\rm\ref{alphaC2}(c)}\,,
\eeq
where ${\cal M}_{\rm\ref{alphaC2}(c)}$ is simply given by vertex
(v) in Eq. (\ref{eftv3}). As a consequence, all diagrams at first
order in $\alpha$ cancel identically and the S-matrix elements
for $2\to 2$ and $3\to 3$ scattering are unchanged by the field
redefinition Eq. (\ref{ftrafo}).

\subsection{Cancellations at Finite Density}

We now proceed to finite density and consider 
a dilute fermion system at zero temperature.
We again concentrate on the case of spin-1/2 fermions. The extension
to include higher spins or isospin is straightforward and would only
change the spin degeneracy factor $g$.
It is not necessary to explicitly introduce a chemical potential
in this case. Note that the number operator changes with
the transformation.  The invariance of the density for a uniform
system is illustrated in the Appendix.

The natural EFT at finite
density is discussed in detail in Ref.~\cite{HAMMER00}. Here, we
only repeat the Feynman rules for the energy density:
\begin{enumerate}
\item Assign nonrelativistic four-momenta (frequency and three-momentum) 
to all lines and enforce four-momentum conservation
at each vertex.
\item 
For each vertex, include the corresponding expression from
Eqs.~(\ref{eftv2}, \ref{eftv3}). 
For spin-independent interactions, the two-body vertices have
the structure $(\delta_{\lambda\gamma}\delta_{\beta\delta}
+\delta_{\lambda\delta}\delta_{\beta\gamma})$,
where $\lambda,\beta$ are the spin indices of the incoming lines
and $\gamma,\delta$ are the spin indices of the outgoing lines.\footnote{
For diagrams with no external legs it is more convenient to take care
of the antisymmetrization in the spin summations. This is
discussed in rule \ref{spinsum}.} For each internal line include a factor 
$iG_0 (\kt)_{\lambda\gamma}$, where
$\kt \equiv (k_0,\vec{k})$ is the four-momentum assigned to the line,
$\lambda$ and $\gamma$ are spin indices, and
\beq
    iG_0 (\kt)_{\lambda\gamma}=i\delta_{\lambda\gamma}
    \left( \frac{\theta(k-\kf)}{k_0-\vec{k}^2/(2M)+i\epsilon}
      +\frac{\theta(\kf-k)}{k_0-\vec{k}^2/(2M)-i\epsilon}\right) \ .
      \label{freeprop_fd}
\eeq
\item Perform the spin summations in the diagram. 
In every closed fermion loop, substitute a spin degeneracy
factor $-g$ for each $\delta_{\lambda\lambda}$.
\label{spinsum}
\item Integrate over all independent momenta with a factor 
$\int\! d^4 k /(2\pi)^4$
where $d^4k\equiv dk_0\,d^3k$. 
If the spatial integrals are divergent, they are defined in $D$ spatial
dimensions and renormalized using minimal subtraction as discussed in
Ref.~\cite{HAMMER00}. For lines originating and ending at the same
vertex (so called tadpoles),
multiply by $\exp(ik_0\eta)$ and take the limit $\eta\to 0^+$
after the contour integrals have been carried out. This procedure
automatically takes into account that such lines must be hole lines.
\item Multiply by a symmetry factor $i/(S \prod_{l=2}^{l_{\rm max}} (l!)^m)$
where $S$ is the number of vertex permutations that transform the diagram into
itself, and $m$ is the number of equivalent $l$-tuples of lines. Equivalent
lines are lines that begin and end at the same vertices with the
same direction of arrows.
\end{enumerate}
\begin{figure}[t]
  \epsfclipon
  \epsfxsize=3.5cm
  \centerline{\epsffile{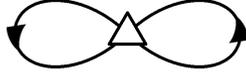}}
\smallskip
\caption{Finite Density: contribution to the energy density
at ${\cal O}(\alpha)$.}
   \label{fdalpha}
\end{figure} 
\noindent
When an off-shell vertex cancels a propagator, a new tadpole can be
generated or a tadpole loop can be shrunk to a point.
We need an additional rule to deal with this situation.
\begin{enumerate}
\item[6.] New tadpoles require a convergence factor as described
in rule 4. Tadpole loops shrunk to a point are zero.
\end{enumerate}

At ${\cal O}(\alpha)$ there is only one diagram, which is 
shown in Fig.~\ref{fdalpha}.
Using the Feynman rules from above and the previous section, we find
\bea
  \edens_5  &=&  g(g-1)\frac{i}{1\cdot2}\lim_{\eta,\eta'\to 0^+}\int
       \frac{d^4 \kt}{(2\pi)^4}\int\frac{d^4 \qt}{(2\pi)^4}e^{i\eta k_0}
       e^{i\eta'\! q_0} iG_0(\kt) iG_0(\qt)\nonumber\\
   & &\times i\frac{4\pi\alpha}{\Lambda^3}\left\{
      2k_0+2q_0-\frac{1}{2M}\left(2\vec{k}^2+2\vec{q}^2\right)\right\}\,.
\eea
After performing the contour integrals, $q_0$ and $k_0$ pick up the poles
from the corresponding propagators, and we find
$\edens_5= 0$. Thus the diagram at ${\cal O}(\alpha )$
vanishes identically, as in the vacuum.

\begin{figure}[t]
   \epsfclipon
   \epsfxsize=11.cm
   \centerline{\epsffile{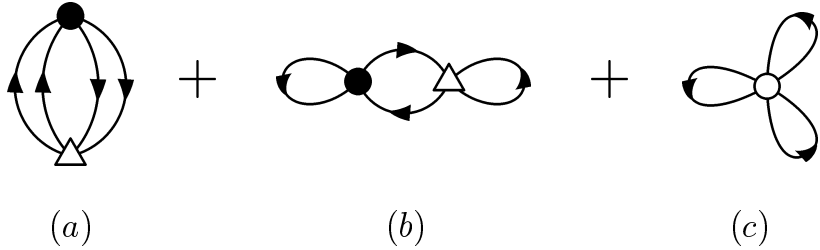}}
\caption{Finite Density: contributions to the energy density 
at ${\cal O}(\alpha C_0)$.}
\label{fdalphaC0}
\end{figure}

We now proceed to ${\cal O}(\alpha C_0)$, where we have the three
diagrams shown in Fig.~\ref{fdalphaC0}.
Using the Feynman rules, we obtain for diagram \ref{fdalphaC0}(a)
\bea
 \edens_{\rm \ref{fdalphaC0}(a)}&=&2g(g-1)\frac{i}{1\cdot 2^2}C_0
     \frac{4\pi\alpha}{\Lambda^3}\int\frac{d^4 \pt}{(2\pi)^4}\int
      \frac{d^4 \kt}{(2\pi)^4}\int\frac{d^4 \qt}{(2\pi)^4}
      G_0(\pt)G_0(\kt)G_0(\pt-\qt)G_0(\kt+\qt)
   \nonumber\\
  & &\times\left\{2k_0+2p_0-\frac{1}{2M}\left(\vec{p}^2+\vec{k}^2+
    (\vec{k}+\vec{q})^2+(\vec{p}-\vec{q})^2\right)\right\}
   \nonumber\\
  &=&
   2g(g-1)i\frac{4\pi\alpha}{\Lambda^3}C_0
              \left[\int\!\!\frac{d^4 \kt}{(2\pi)^4}G_0(\kt)
   \right]^3=2g(g-1)\frac{4\pi\alpha}{\Lambda^3}C_0\frac{\kf^9}{216\pi^6}\,,
\eea
where the off-shell vertex was canceled against the propagators with
appropriate momenta. Interestingly,
the diagram \ref{fdalphaC0}(b) does not vanish separately even at $T=0$, 
as one would  naively expect. The reason is that the off-shell vertex cancels
one of the propagators in the middle loop with two propagators.
Consequently, the product $\theta(p-\kf)\theta(\kf-p)$
corresponding to a particle-hole pair with equal momentum,
which usually makes such diagrams vanish at $T=0$, does not arise. 
In particular, we have
\bea
 \edens_{\rm \ref{fdalphaC0}(b)}&=&-g(g-1)^2\frac{i}{1\cdot 1}C_0
    \frac{4\pi\alpha}{\Lambda^3}\lim_{\eta,\eta'\to 0^+}
       \int\frac{d^4 \kt}{(2\pi)^4}
    \int\frac{d^4 \qt}{(2\pi)^4}\int\frac{d^4 \pt}{(2\pi)^4}e^{i\eta k_0}
       e^{i\eta' q_0} G_0(\kt) G_0(\qt) \left(G_0(\pt)\right)^2
      \nonumber\\
  & &\times\left\{2k_0+2p_0-\frac{1}{2M}\left(2\vec{p}^2+2\vec{k}^2\right)
      \right\}
      \nonumber\\
  &=&-2g(g-1)^2 \frac{4\pi\alpha}{\Lambda^3}C_0\frac{\kf^9}{216\pi^6}\,.
\eea
Finally, we obtain for diagram \ref{fdalphaC0}(c)
\bea
  \edens_{\rm \ref{fdalphaC0}(c)}&=&-g(g-1)(g-2)\frac{i}{1\cdot 3!}(-i)12 C_0
     \frac{4\pi\alpha}{\Lambda^3} \left[\int\!\!\frac{d^4 \kt}{(2\pi)^4}
      iG_0(\kt) \right]^3
   \nonumber\\
 &=& 2g(g-1)(g-2)\frac{4\pi\alpha}{\Lambda^3}C_0\frac{\kf^9}{216\pi^6}\,.
 \label{eq:falphaC0}
\eea
Adding all contributions to the energy density at ${\cal O}(\alpha C_0)$,
the sum of $\edens_{\rm \ref{fdalphaC0}(a)}$ and  
$\edens_{\rm \ref{fdalphaC0}(b)}$ builds up exactly the same spin 
degeneracy factor as occurs in $\edens_{\rm \ref{fdalphaC0}(c)}$, and
we have
\beq
  \edens_{\rm \ref{fdalphaC0}(a)} + \edens_{\rm \ref{fdalphaC0}(b)}
        + \edens_{\rm \ref{fdalphaC0}(c)}=0\,.
\eeq
\begin{figure}[t]
   \epsfclipon
   \epsfxsize=11.cm
   \centerline{\epsffile{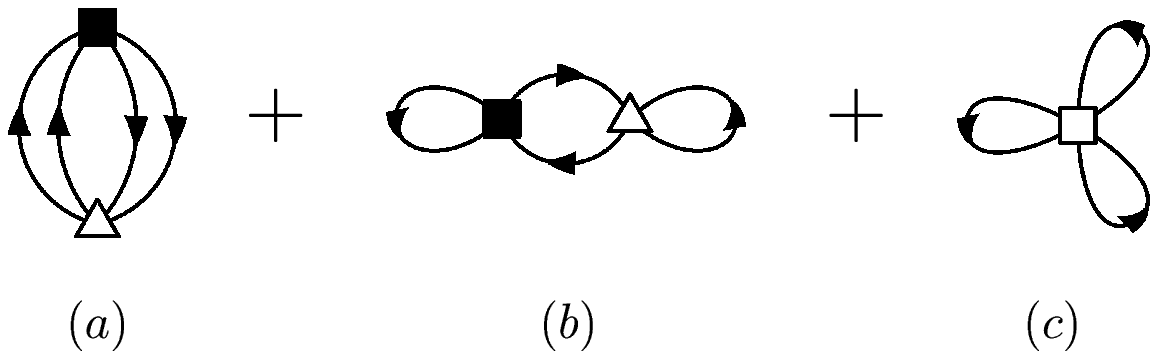}}
\caption{Finite Density: contributions to the energy density 
at ${\cal O}(\alpha C_2)$.}
\label{fdalphaC2}
\end{figure} 
The corresponding diagrams at ${\cal O}(\alpha C_2)$
are shown in Fig.~\ref{fdalphaC2}. After their evaluation, we find
\bea
  \edens_{\rm \ref{fdalphaC2}(a)} &=& g(g-1)\frac{4\pi\alpha}{\Lambda^3}
  C_2\frac{\kf^{11}}{180\pi^6}\,,\nonumber\\
 \edens_{\rm \ref{fdalphaC2}(b)} &=& -\frac{1}{2}g(g-1)^2
   \frac{4\pi\alpha}{\Lambda^3}C_2\frac{\kf^{11}}{180\pi^6}\,,\nonumber\\
 \edens_{\rm \ref{fdalphaC2}(c)} &=& \frac{1}{2}g(g-1)(g-3)
    \frac{4\pi\alpha}{\Lambda^3}C_2\frac{\kf^{11}}{180\pi^6}\,,
\eea
where the sum of $\edens_{\rm \ref{fdalphaC2}(a)}$ and 
$\edens_{\rm \ref{fdalphaC2}(b)}$ again builds up the same 
spin degeneracy factor as in $\edens_{\rm \ref{fdalphaC2}(c)}$
and the net contribution of all diagrams at ${\cal O}(\alpha C_2)$
is zero. In contrast to the contribution at ${\cal O}(\alpha C_0)$,
the induced three-body term with derivatives generates a factor of $g-3$ 
instead of $g-2$ and does not vanish separately for spin-1/2 fermions.
In conclusion, the contribution of all first order diagrams
in $\alpha$ vanishes exactly, and the field transformation Eq. (\ref{ftrafo})
does not change the energy density.

\begin{figure}[t]
   \epsfclipon
   \epsfxsize=11.cm
   \centerline{\epsffile{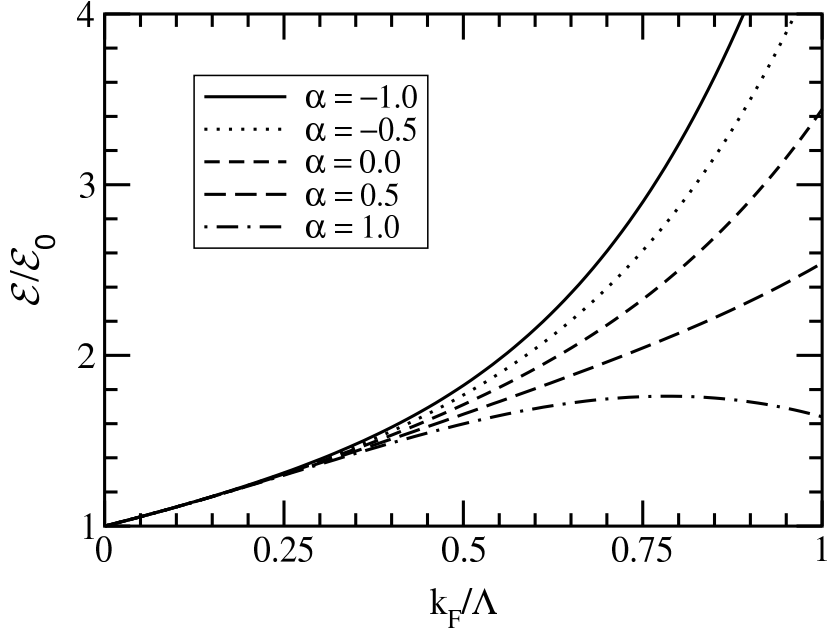}}
\smallskip
\caption{Energy density for a dilute Fermi system with spin
degeneracy $g=4$ through ${\cal O}(\kf^9)$
as a function of $\kf/\Lambda$, normalized to the energy density
${\cal E}_0$ for a free Fermi gas. 
The induced three-body contribution Eq.~(\ref{eq:falphaC0}) is
omitted for $\alpha\neq0$.
When it is included, all curves coincide with the $\alpha=0$ curve.}
\label{fdcoester}
\end{figure} 

These cancellations manifest the invariance of finite density observables
under field redefinitions.
It is evident that one must include all induced terms.
However, in past investigations with unitarily equivalent two-body
potentials, possible induced three-body contributions were not considered.
(As demonstrated in the next section, unitary transformations of the
Hamiltonian have the same effect as field redefinitions.)
Fig.~\ref{fdcoester} illustrates the consequence of omitting these
terms for a dilute Fermi gas with spin degeneracy $g=4$.
The energy density of the gas through ${\cal O}(\kf^9)$
(see \cite{HAMMER00} for details) is plotted for different values of
$\alpha$, with the contributions ${\cal E}_{6(a)}$ and
${\cal E}_{6(b)}$ with two-body vertices included but the 
contribution ${\cal E}_{6(c)}$ with a three-body vertex not included.
The result is a band of energies, analogous to the ``Coester band''
found for nuclear matter saturation curves \cite{COESTER70}.
The normalization of the transformation was chosen so that 
$\alpha=1$ is natural-sized; a manifestation is that the ``correction''
for $\kf=\Lambda$ is order unity for $\alpha=1$.
The spread of the Coester line curves in the medium
is a measure of the natural size of three-body contributions. 

In summary,
we have used a simple field redefinition to illustrate how two-body
off-shell vertices are induced but canceled exactly by new many-body
vertices.
Although we have worked to leading order in the EFT perturbation theory
(that is, to leading order in a momentum expansion),
the cancellation is guaranteed order-by-order.\footnote{As noted earlier,
a similar cancellation holds in a finite system with a single
particle potential included in the noninteracting propagator.}
Next we show how unitary transformations on NN potentials exhibit the
same behavior. 


\section{Unitary Transformations and Field Redefinitions}    
\label{unitary}

Infinitely many potentials that predict the
same scattering phase shifts and energy spectrum can be constructed
using unitary transformations \cite{EKSTEIN60}.
If
\beq
  H \Psi = (H_0 + V) \Psi = E \Psi
    \label{eq:H}
\eeq
and $\exp(i S)$ is a unitary operator of finite range:
\beq
  e^{iS} \Psi = \widetilde\Psi 
      \stackrel{\mathstrut r\rightarrow\infty}{\longrightarrow}
       \Psi
       \ ,  \label{eq:U}
\eeq
then the transformed Hamiltonian
\beq
   \widetilde H = e^{iS} H e^{-iS} \equiv H_0 + \widetilde V
     \label{eq:Htilde}
\eeq
has the same energy spectrum and phase shifts at all energies as $H$.
The potentials $V$ and $\widetilde V$ are said to be phase-shift equivalent.
Various practical methods to construct families of such potentials are
described in the literature 
\cite{COESTER70,SRIVASTAVA70,HAFTEL71,MONAHAN71,EKSTEIN60}.

Here we demonstrate that these short-range
unitary transformations have the same qualitative behavior as the simple
field redefinitions in the last section: off-shell contributions are generated
at the same time as compensating many-body contributions.
This is only evident upon embedding the problem in a many-body framework
from the beginning.  
The interplay between off-shell behavior and many-body forces
in the context of unitary transformations
has been studied in the literature \cite{POLYZOU90,AMGHAR95}.
In particular,
Ref.~\cite{POLYZOU90} proves a general theorem that guarantees the existence
of three-body potentials that, when combined with unitarily equivalent
two-body potentials, yield identical three-body observables.

We adapt the discussion of unitary transformations in Ref.~\cite{AMGHAR95},
which addresses the key points (although we will use the notation
of Ref.~\cite{FETTER71}).
By writing the Hamiltonian and unitary operator in second-quantized
form, the transformation becomes a canonical transformation of the
creation and destruction operators.
We consider spin-0 boson operators for simplicity; generalizing
to spin-1/2 fermions is direct but would obscure the main points.
We  focus on a special case that highlights the relationship
to local field redefinitions.

A Hamiltonian with kinetic energy and a
two-body potential $V$ is written in terms of
second-quantized field operators as \cite{FETTER71}
\beq
\wh H =
  \int\! d^3x\, \psihat^\dagger(\xvec)\frac{-\nabsq}{2M}
  \psihat(\xvec)
  + \frac{1}{2}  \int\!\!\int\! d^3x\, d^3x'\,
  \psihat^\dagger(\xvec) \psihat^\dagger(\xvec')
  V(\xvec,\xvec') \psihat(\xvec')  \psihat(\xvec)
  \ .
\eeq 
Amghar and Desplanques \cite{AMGHAR95} 
consider unitary transformations of the form
\beq
  \wh H \longrightarrow e^{i{\wh S}} \wh H e^{-i{\wh S}}
        = \wh H + i[\wh S,\wh H] + \cdots
        \label{eq:Htrans}
\eeq
with 
\beq
  \wh S \equiv  \int\!\!\int\! d^3x\, d^3x'\,
  \psihat^\dagger(\xvec) \psihat^\dagger(\xvec')
  U(\xvec-\xvec') \psihat(\xvec')  \psihat(\xvec)
  \label{eq:Sdef}
\eeq
and $U$ Hermitian.
The simplest nontrivial form that maintains time-reversal 
invariance \cite{AMGHAR95}
is
\beq
  U(\xvec) = \{ {\bf p},\xvec f(x) \} = i\nabl_{\!\xvec}\, \xvec f(x)+ 
    \xvec f(x) (-i)\nab_{\!\xvec}\,
  \label{eq:Uform}
\eeq
where ${\bf p}$ is the momentum operator and the curly brackets denote
an anti-commutator.
They show that the transformation (\ref{eq:Htrans})--(\ref{eq:Uform})
generates two-body off-shell contributions but
also many-body terms.

This transformation is nonlocal for a finite range $U$, which complicates
the manipulations.  
If the nonlocality is of the
order of the short-distance scale $\Lambda$, from an EFT perspective
it is appropriate to replace $U(\xvec)$ with contact terms.
That is, $U(\xvec)$ becomes proportional to a delta function with 
derivatives. The simplest choice is 
\beq
\xvec f(x) = \alpha \nab\delta^3(\xvec)\ .
\label{eq:fxform}
\eeq
where $\alpha$ is an arbitrary parameter.
Since the interaction is singular,
we can expect to generate divergent contributions, which we will
regulate with dimensional regularization.

The transformation is most transparent in momentum space, with
field expansions (in a box of volume $V$ with 
periodic boundary conditions) \cite{FETTER71}:
\beq
  \label{eq:fexp}
  \psihat(\xvec) = \frac{1}{\sqrt{V}}\sum_\kvec e^{i\kvec\cdot\xvec} \ak \ ,
  \qquad
  \psihat^\dagger(\xvec) =
     \frac{1}{\sqrt{V}}\sum_\kvec e^{-i\kvec\cdot\xvec} \akdag\ ,
\eeq
and boson commutation relations for the operators $\ak$ and $\akdag$:
\beq
    \label{eq:comm}
    [\ak,\akpdag] = \delta_{\kvec\kvec'} \ ,
    \qquad
    [\ak,\akp] = [\akdag,\akpdag] = 0 \ .
\eeq
The Hamiltonian for a zero-range potential is
\beq
  \widehat H = 
  \sum_{\kvec} \frac{\kvec^2}{2M}\, \akdag\ak 
     + \frac{C_0}{2V}
     \sum_{\kvec_1,\kvec_2,\kvec_3,\kvec_4}
     \akdagv1 \akdagv2 \akv3 \akv4
     \delta^\dagphan_{\kvec_1+\kvec_2,\kvec_3+\kvec_4}
     \ ,
\eeq
with $C_0$  given in Eq.~(\ref{C2imatch}).
This Hamiltonian corresponds to the Lagrangian of Eq.~(\ref{lag})
with the $C_2$ term omitted.
In order to obtain the transformation operator $\widehat{S}$,
we substitute Eqs.~(\ref{eq:Uform}) and  (\ref{eq:fxform}) into 
Eq.~(\ref{eq:Sdef}). After changing the integrations over $x$ and $x'$
into integrations over $z_\pm=x\pm x'$, the $z_+$ integration
simply gives  overall momentum conservation. The remaining
integral can be carried out using a partial integration, and the 
transformation operator becomes
\beq
  \wh S = i\frac{\alpha}{4V}
          \sum_{\kvec_1,\kvec_2,\kvec_3,\kvec_4}
        \Bigl( (\kvec_1-\kvec_2)^2 - (\kvec_3-\kvec_4)^2
        \Bigr)\,
     \akdagv1 \akdagv2 \akv3 \akv4
     \delta^\dagphan_{\kvec_1+\kvec_2,\kvec_3+\kvec_4} \ .
\eeq
The transformed Hamiltonian is
\bea
  \widehat H' &=& \wh H +  i[\wh S, \wh H] +
         \order{\alpha^2}
       \nonumber  \\
       &=&  \wh H
  + \frac{\alpha}{16MV}     
          \sum_{\kvec_1,\kvec_2,\kvec_3,\kvec_4}
        \Bigl( (\kvec_1-\kvec_2)^2 - (\kvec_3-\kvec_4)^2
        \Bigr)^2\,
     \akdagv1 \akdagv2 \akv3 \akv4\,
     \delta^\dagphan_{\kvec_1+\kvec_2,\kvec_3+\kvec_4} 
     \nonumber \\[5pt]
     & & -\frac{\alpha C_0}{2V^2}  
         \sum_{\kvec_1,\kvec_2,\kvec_3,\kvec_4,\kvec_5} 
         \Bigl( (\kvec_1-\kvec_2)^2 - (2\kvec_5-\kvec_3-\kvec_4)^2
        \Bigr)\, \akdagv1 \akdagv2 \akv3 \akv4\,
        \delta^\dagphan_{\kvec_1+\kvec_2,\kvec_3+\kvec_4}  
     \nonumber \\
     & &-\frac{\alpha C_0}{2V^2}
         \sum_{\kvec_1,\kvec_2,\kvec_3,\kvec_4,\kvec_5,\kvec_6} 
         \Bigl( (\kvec_1-\kvec_2)^2 - (\kvec_4-\kvec_5-\kvec_6+\kvec_3)^2
                +(\kvec_4-\kvec_5)^2 
     \nonumber \\
      & &  \quad - (\kvec_1-\kvec_2-\kvec_3+\kvec_6)^2
         \Bigr)\,  \akdagv1 \akdagv2 \akdagv3 \akv4 \akv5 \akv6\,
         \delta^\dagphan_{\kvec_1+\kvec_2+\kvec_3,\kvec_4+\kvec_5+\kvec_6}  
       +  \order{\alpha^2}
\label{eq:Htransformed}
\eea
The two-body term generated from the commutator with the potential
contains a divergent sum over the momentum $\kvec_5$.  
However, this is a
pure power divergence, which is zero in dimensional regularization
(with a cut-off regulator, it would simply renormalize the $C_0$
coefficient in the effective Lagrangian).

We observe that the momentum dependence of the first two-body term,
which results from the commutator with the kinetic term, can be
written
\beq
   (\kvec_1-\kvec_2)^2 - (\kvec_3-\kvec_4)^2
     = 2(\kvec_1^2 + \kvec_2^2 - \kvec_3^2 - \kvec_4^2) \ .
\eeq
This factor vanishes for on-shell legs and so does not contribute to
two-body scattering.
The analysis proceeds in parallel with the analysis
in the last section; we find the
analog of
Fig.~\ref{alphaC0} with the same result:  the off-shell and three-body
contributions cancel for $3\to 3$ scattering or in the medium.

We can identify a local field redefinition equivalent to this unitary
transformation by directly transforming the field operator:
\beq
  \psihat(\xvec) \longrightarrow e^{i\wh S} \psihat(\xvec) 
             e^{-i\wh S}
         =   \psihat(\xvec) + i [\wh S, \psihat(\xvec)] +
         \order{\alpha^2} \ .
         \label{eq:psitran}
\eeq
The commutator is conveniently evaluated by expanding in the
momentum basis, Eq.~(\ref{eq:fexp}), and using the commutation relations,
Eq.~(\ref{eq:comm}).
The end result can then be interpreted as a field redefinition in
the effective Lagrangian:
\beq
 \psi \longrightarrow \psi +\alpha 
   \Bigl[
     (\nabsq\psi^\dagger)\psi\psi
     + \psi^\dagger(\nab\psi)^2
     + (\nab\psi^\dagger)\cdot\nab(\psi\psi)
   \Bigr]
    +  \order{\alpha^2} \ ,
\eeq
with the corresponding redefinition of $\psid$.
This transformation generates a vertex equivalent to the term
proportional to $\alpha/(MV)$ in 
Eq.~(\ref{eq:Htransformed}) but also appears to generate a vertex with 
time-derivatives from the term,
\beq
 \Delta{\cal L} = \alpha \Bigl[\psi^\dagger \psi^\dagger (\nabsq\psi) 
     + (\nab\psi^\dagger)^2 \psi \Bigr](i\dt\psi) + \mbox{H.c.} \ ,
\eeq
which does not appear in the transformed Hamiltonian.  
However, the Feynman rule for this
vertex is proportional to $\omega_1+\omega_2-\omega_3-\omega_4$,
which vanishes identically by energy conservation.

In Ref.~\cite{FST00}, EFT at finite density was explored by studying
a perturbative matching example.  
In particular, a model potential
\beq
	\langle {\bf k'}|\hat V_{\rm true}|{\bf k}\rangle = 
		\frac{4\pi}{M}  \frac{\gamma m^3}{(k^2+m^2)(k'{}^2+m^2)} \ .
	\label{true}
\eeq
was used to represent the underlying dynamics, where the 
mass $m$ corresponds to the range (and non-locality) of the potential;
it plays the role of the underlying short-distance scale $\Lambda$.
The dimensionless coupling $\gamma$ provides a perturbative
expansion parameter.
An EFT was introduced to reproduce observables in a double expansion:
order-by-order in $\gamma$ as well as the EFT momentum expansion.

An effective $s$-wave potential in the form
\beq
  \langle {\bf k'}|\hat V_{\rm EFT}|{\bf k}\rangle
     = C_0 + C_2 \frac{\left( k^2+k'{}^2 \right)}{2} +
    C_4 \frac{\left( k^2 + k'{}^2 \right)^2}{4} +
	\wt C_4 \frac{\left(k^2-k'{}^2 \right)^2}{4} + \ldots \ ,
 \label{veff}
\eeq
was used. The EFT expansion of this potential was studied using 
both dimensional regularization and cut-off regularization.
The term proportional to $\wt C_4$ does not contribute to on-shell
two-body scattering in dimensional regularization, but
is essential in the matching of the energy per particle in 
an infinite system.
The value of $\wt C_4$ was determined by matching on-shell $3\to 3$
scattering in Ref.~\cite{FST00}.
But it is tempting to conclude that this is merely equivalent to
matching the T-matrices generated by Eqs.~(\ref{true}) and (\ref{veff})
{\em off shell\/}.

However, the unitary transformation discussed above shows that we can
arbitrarily
trade-off the value of $\wt C_4$ with the coefficient of a three-body 
contact term. 
This means that  off-shell matrix elements of the potential (e.g.,
$\langle {\bf 0}|\hat V_{\rm EFT}|{\bf k}\rangle$) and subsequently
the off-shell T-matrix can be changed continuously by varying $\alpha$,
without changing {\em any\/} observables. 
One choice is to eliminate two-body off-shell vertices such as $\wt C_4$
entirely in favor of many-body vertices that do not vanish on shell. 
This is the form of many-body EFT used in Ref.~\cite{HAMMER00},
which corresponds to Georgi's ``on-shell effective field theory''
\cite{GEORGI91}.
In different situations, different choices may be more efficient
\cite{CRS99}.


\section{Summary}
\label{conclusions}

The attitude in the mid-seventies 
about off-shell physics
is summarized in the review article
by Srivastava and Sprung \cite{SRIVASTAVA75}:
\begin{quote}
 ``A knowledge of the off-energy-shell behavior of the nucleon-nucleon
 interaction is basic to nuclear physics.  The on-energy-shell information,
 however complete (which it is not), is not adequate to permit unambiguous
 calculation of the properties of systems of more than two nucleons.
 Any reasonable agreement with nuclear binding energy and other properties
 obtained by using different potential models without knowing the correct
 off-energy-shell behavior of the interaction is probably fortuitous
 and therefore not very meaningful.  The results obtained with different
 potentials only rarely agree with each other.''
\end{quote}
In contrast, effective field theories {\em are\/} determined
completely by on-energy-shell information, up to a well-defined
truncation error. In writing down the most general Lagrangian 
consistent with the symmetries of the underlying theory,
many-body forces arise naturally. Even though they are 
usually suppressed at low energies, they enter at some order
in the EFT expansion. These many-body forces have to be determined
from many-body data. The key point is, however, that no
off-energy-shell information is needed or experimentally accessible.

In principle, one could consider the formal problem
of particles interacting via a definite two-body potential, valid for
all energies. From a physical point of view, such a description has
to break down at a certain energy, when, for example, the substructure
of the particles can be resolved. Furthermore, such a scenario cannot
correspond to QCD where three-body forces are present at a fundamental
level. The EFT approach is ideally suited to deal with such a situation,
because at low energies all sensitivity to the high-energy dynamics
can be captured in a few low-energy constants.

The variation in observables calculated using the 
two-body parts only when short-range unitary transformations are 
performed gives an indication of 
the natural size of three-body contributions. Given the correspondence
of unitary transformations and field redefinitions (as shown in the
previous section), this statement is a simple consequence of naive 
dimensional analysis in the EFT. In cases where naive dimensional analysis
fails, further investigation is needed.

The modern consensus is that at least 
three-body forces are required as supplements to the best phenomenological
NN interactions; discrepancies between calculated and experimental 
binding energies for
the triton have made this conclusion unavoidable.  
But the emphasis
on the importance of reproducing the ``correct'' off-shell physics 
still persists.  
For example,
in a recent review of the state-of-the-art Bonn nucleon-nucleon
potential, differences in the off-shell behavior of the
Bonn potential compared to other contemporary NN potentials are
stressed \cite{MACHLEIDT00}.
The hope of using many-nucleon systems to decide on the ``correct''
off-shell potential also persists \cite{POLLS98,DOLESCH00}.
Again, the EFT viewpoint is that all theories with consistent power
counting {\em must\/} agree up to truncation errors.

We have noted several
consequences of the freedom to perform field redefinitions:
\begin{itemize}
  \item An effective Lagrangian with 
  energy-independent effective potentials can always be used, by eliminating
  time derivatives with field redefinitions.
  However, it is sometimes numerically convenient to trade momentum
      dependence for energy dependence \cite{LEPAGE89}, or to retain
      time dependence when incorporating relativistic corrections
      \cite{CRS99}. 
  \item The Lagrangian can be restricted to on-shell vertices only
  (cf.\ Georgi's ``on-shell effective field theory'' \cite{GEORGI91}).
    One advantage of this choice is that the maximal number of energy diagrams
    vanish (at zero temperature).
  \item One can choose to trade off-shell contributions for
  many-body forces.
  If minimizing many-body forces is desirable, then one could argue
  that adjusting off-shell behavior to achieve this end is optimal.
  However, fine tuning is required to reduce many-body contributions
  below their ``natural'' size.
\end{itemize}
Finally, we
note that different {\em regularizations\/} will also yield very different
potentials, with different off-shell behavior, but observables will agree. 

\acknowledgements
We acknowledge useful discussions with G.~P.\ Lepage, T.~Mehen,
  R.~J.\ Perry, M.~Savage, B.~D.\ Serot, 
  J.~V.\ Steele, U. van Kolck, and M.~Wise.
RJF and HWH thank the Institute for Nuclear Theory at the University
of Washington for its hospitality and partial support during completion
of this work.
This work was supported by the National Science Foundation
under Grant No.\ PHY--9800964.

\begin{appendix}
\section*{Number Operator}
\begin{figure}[t]
  \epsfclipon
  \epsfxsize=13cm
  \centerline{\epsffile{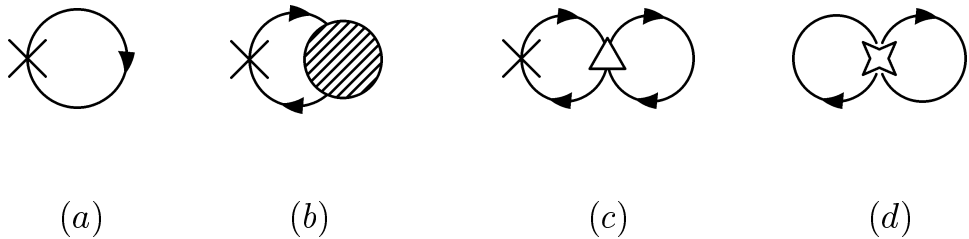}}
\smallskip
\caption{Finite Density: diagrams for the number operator.}
   \label{fdnumber}
\end{figure}

The field transformation, Eq.~(\ref{ftrafo}), changes the 
number operator of the theory, which is defined as the time component
of the Noether current associated with the phase transformation
$\psi \to \exp(-i\phi)\psi$. In the untransformed theory,
the number operator is simply $\wh{N}=\psid\psi$. 
The number density $n_0$ for the free Fermi gas at $T=0$ is given by the 
Feynman diagram shown in Fig.~\ref{fdnumber}(a), where the cross
indicates the insertion of the number operator.
Evaluation of the Feynman diagram gives
\beq
n_0=-g \lim_{\eta\to 0^+}\int\frac{d^4 \pt}{(2\pi)^4} iG_0(\pt)e^{i\eta p_0}
=g\frac{\kf^3}{6\pi^2}\,.
\eeq
This density $n_0$ remains unchanged by interactions
and $n\equiv n_0$, which can be 
understood as follows. All contributions from interactions 
have the form shown in Fig.~\ref{fdnumber}(b), where the blob represents 
the remainder of a given Feynman diagram. The crucial point is that there
are always two different propagators with the same momentum connected
to the insertion of $\wh N$. As a consequence, there is no simple 
pole in the time component of the corresponding momentum, and the contour 
integral over the time component vanishes. This argument is valid as
long as there are no off-shell terms in the vertices that cancel one
of the propagators in question.

After the field transformation, Eq.~(\ref{ftrafo}), the number operator 
acquires an additional term of ${\cal O}(\alpha)$ and becomes
\beq
\wh{N}_\alpha = \psid\psi +\frac{4\pi\alpha}{\Lambda^3}2\left(\psid\psi
\right)^2\,.
\eeq
The expectation value of the number operator and the density, however, 
remain unchanged. We show this explicitly
for the lowest order in $\alpha$. At order $\alpha$, we have 
only the two new Feynman diagrams shown in Figs.~\ref{fdnumber}(c) and
\ref{fdnumber}(d). The cross indicates an insertion of the part of the number 
operator that is ${\cal O}(1)$, while the tetragram denotes
the part of the number operator that is  ${\cal O}(\alpha)$.
The triangle denotes the induced off-shell vertex (iii) 
(cf. Eq. (\ref{eftv2})). The diagram 
\ref{fdnumber}(c) does not vanish from the general argument above
because the off-shell vertex cancels one of the propagators connected
to the insertion of $\wh N$. Evaluating the two diagrams, we find
\begin{eqnarray}
n_{\ref{fdnumber}(c)}
&=&-g(g-1)\frac{4\pi\alpha}{\Lambda^3}\frac{\kf^6}{36\pi^4}
=-n_{\ref{fdnumber}(d)}\ .
\end{eqnarray}
Thus the contributions cancel and the density is unchanged 
to ${\cal O}(\alpha)$.
Higher order contributions in $\alpha$ cancel in a similar way.
\end{appendix}


\end{document}